# Superconductivity in $Hf_5Sb_{3-x}Ru_x$: Are Ru and Sb a critical charge-transfer pair for superconductivity?

Weiwei Xie*[1], Huixia Luo[1], Elizabeth M. Seibel[1], Morten B. Nielsen[2] and Robert J. Cava*[1]

[1] Department of Chemistry, Princeton University, Princeton, NJ, USA, 08540
[2] Center for Materials Crystallography (CMC), Department of Chemistry and iNANO, Aarhus University, 8000 Aarhus C, Denmark,

*Abstract*

We address the hypothesis that in intermetallic compounds, contrary to a long-standing view that considers electron count and crystal structure type as the only significant chemical criteria for the occurrence of superconductivity, consideration of the actual elements present is a third equally important factor. The importance of chemical identity may seem obvious, especially to chemists, but it has not previously been explicitly tested for intermetallic superconductors. Here we test the hypothesis by searching for and finding a new superconductor in the tetragonal symmetry $Hf_5Sb_{3-x}M_x$ solid solution. This phase is ideal for this study, because M can be many $3d$, $4d$ and $5d$ transition metals in an M-Sb chain and is a minor elemental constituent. We find superconductivity for M = Ru only. This is the case even when the electron count can be adjusted to the same value with a different transition element within the same structure type. This leads us to propose that, like Cu-O (cuprate superconductors) and Fe-As (iron-pnictide superconductors) in different classes of compounds, Ru and Sb may be a critical element pair for superconductivity in intermetallic phases.

Corresponding author: rcava@princeton.edu (Prof. Robert Cava);
weiweix@princeton.edu (Dr. Weiwei Xie)

Superconductivity remains unpredictable for new compounds because, from the physics perspective, it results from instabilities in a compound's electronic system that are delicately balanced with other factors such as electron-lattice coupling or magnetism.[1] As such, there are few predictive rules that are thought to work more than they fail. One of these; perhaps the most long-standing, is that within intermetallic compounds of a known superconducting structure type, one can count electrons and expect to find the best superconductivity at 4.5 or 6.5 valence electrons per atom.[2, 3] This perspective can be considered a physicist's "rigid band" model where the structure type and electron count are held paramount; the actual atoms present are generally not a prime consideration.

But, from a chemical perspective, there have long been indications that there is much more to the presence or absence of superconductivity than just electron count and structure type. Simply put, there appear to be a handful of "critical pairs" of atoms in the periodic table, where it can be postulated that the balance between covalent and ionic bonding leads to just the right kind of charge transfer between the atoms so that the bond valence responds to perturbations from the other forces present to lead to a superconductor. In contrast to the quantitative $k$ space or Fermi surface view often employed in the hunt for superconductivity, the concept of critical charge transfer pairs is clearly a qualitative, real space view of what can give rise to superconductivity, however, this can be an equally important and productive viewpoint, especially from the perspective of the synthetic chemistry. [4] Many would agree, for example, that clear examples of such critical charge-transfer pairs in the periodic table are Cu-O and Fe-As, with other possibilities hinted at as well.[5, 6, 7]

We were therefore motivated to test the importance of chemical identity on superconductivity in intermetallic compounds. We began by looking at compounds of the type $(Hf/Zr)_5(Sb_{3-x}M_x)$ where $M$ is a transition metal and $x \sim 0.5$, which occur within the superconducting $W_5Si_3$ structure type ($W_5Si_3$, $T_c$ = 2.7 K). Importantly, the doping of M stabilizes this structure type and allows us to elucidate the effect of element identity only. Despite the stabilization of a superconducting structure type, no superconductivity had been found in such systems prior to this study. As shown in Figure 1(Upper), $M$ and Sb form a bonded chain in these compounds, inside a column of Hf. The $M$ atom is a small fraction of the atoms present, and so one might initially guess that it is a trivial contributor to the overall electronic structure of the compound.

However, our electronic structure calculations, described below, indicate otherwise. Additionally, the calculations show the presence of a saddle point in the energy vs. wavevector dispersion of one of the electronic states, which is a characteristic of band structures often taken to be a sign of impending electronic instability and a propensity toward superconductivity.[8] Yet another factor in the recipe for superconductivity is often the suppression of magnetism. Ferromagnetism has been reported for $Hf_5Sb_{3-x}M_x$ when $M$ = Fe, which implies that the $3d$ electrons may not be fully hybridized with the near neighbor Sb atoms and that this compound may sit near a critical atom-pair boundary. In such cases, the replacement of such a $3d$ element for a $4d$ element can give rise to superconductivity because $d$ electrons in $4d$ and $5d$ transition metals may hybridize with neighboring Sb atoms. This motivated us to search for stable compounds in the same structure type based on $4d$ elements rather than $3d$ elements. Naturally therefore, our primary candidate for superconductivity was $M$ = Ru.

Indeed, we find that the Ru variant $Hf_5Sb_{2.5}Ru_{0.5}$ is superconducting at 3.2 K; however, we do not find any superconductivity present in the same structure type when Mo, Rh, Pd, or Pt is present in place of Ru, nor have others found it when the $3d$ elements V, Mn, Co, Fe, Ni, and Cu are present.[9] Thus of 11 transition elements tested, superconductivity is present for Ru only. Critically, we also do not find superconductivity when different atoms are co-substituted as minor constituents to yield an isoelectronic band structure to $Hf_5Sb_{2.5}Ru_{0.5}$; $(Hf_{4.5}Y_{0.5})Sb_{2.5}Rh_{0.5}$, for example, which is isostructural and isoelectronic with our superconductor, $Hf_5Sb_{2.5}Ru_{0.5}$, is not superconducting. These clear observations, along with the fact that the skutterudite $LaRu_4Sb_{12}$ is superconducting while isoelectronic and isostructural $LaRh_4Sb_8Sn_4$ is not,[10,11] lead us to propose that Ru-Sb may be a third critical charge-transfer pair of elements for superconductivity in the periodic table along with Cu-O and Fe-As. Our work is described in detail in the following.

Polycrystalline $Hf_5Sb_{3-x}Ru_x$ samples were synthesized by arc melting the elements in a water-cooled copper hearth under an argon atmosphere using a tungsten electrode. The starting materials, hafnium (powder, 99.9%, Alfa Aesar), ruthenium (powder, 99.95%, Aldrich), and antimony (crystalline pieces, 99.999%, J&M) were weighed in the $Hf_5Sb_{3.0-x}Ru_x$ ($x$ = 0.0, 0.2, 0.4, 0.6, 0.8 and 1.0) stoichiometric ratios (total mass 300 mg; 10% molar excess Sb added in order to balance Sb loss during the arc melting), pressed into pellets, and arc melted for 10 seconds.

The products were turned and melted several times to ensure good homogeneity. Weight losses during the melting process were less than 2%. The products were stable in air and moisture. $Hf_5Sb_{3-x}Ru_x$ with $x = 0.0$ and $0.2$ crystallizes in the $Y_5Bi_3$-type structure; these materials are not superconducting and are not studied further here. With $x$ in the range from 0.4 to 0.8, the product adopts the $W_5Si_3$-type structure, the subject of this work, whereas at higher doping levels ($x = 1.0$), an impurity phase, HfRu, appears in the final products. The as-melted $Hf_5Sb_{3-x}Ru_x$ samples were examined by powder X-ray diffraction for identification and phase purity on a Bruker D8 ECO powder diffractometer employing Cu Kα radiation with aid of a full-profile Rietveld refinement using Fullprof.[12,13] The phase in the powder pattern is a good fit to the $Hf_5Sb_{2.54(4)}Ru_{0.46}$ structural model obtained from the single crystal study. The quantitative analysis of the powder diffraction pattern showed that the polycrystalline sample employed for the bulk property characterization consists of pure $Hf_5Sb_{3-x}Ru_x$ (see Supplementary Information Figure 1(bottom)). For the purposes of property comparison, $Hf_5Sb_{2.5}M_{0.5}$ ($M$ = Mo, Rh, Pd, Re, Ir and Pt), $Hf_{4.5}Y_{0.5}Sb_{2.5}Rh_{0.5}$ and HfRu were prepared as pure phases by arc melting the elements in a stoichiometric ratio.

To specify the structure of the $Hf_5Sb_{3-x}Ru_x$ compound, single crystals extracted from arc melted samples were investigated on a Bruker Apex Phonon diffractometer with Mo radiation Kα$_1$ (λ=0.71073 Å). The crystal structure was solved using direct methods and refined by full-matrix least-squares on $F^2$ with the SHELXT package,[14,15] and the chemical compositions of the $Hf_5Sb_{2.54(4)}Ru_{0.46}$ (hereafter referred to as $Hf_5Sb_{2.5}Ru_{0.5}$) crystals from the as-cast samples studied were confirmed by SEM-EDX chemical analysis performed on a Quanta 200 FEG ESEM operated at 20kV. Temperature ($T$) dependent electrical resistivity (ρ) was measured from 1.9 and 300 K with the four-probe technique using silver paste electrodes on a Quantum Design Physical Property Measurement System (PPMS). Zero-field cooled (ZFC) Magnetic susceptibility ($\chi_{mol}(T)$), measured in a field of 10 Oe using a Quantum Design superconducting quantum interference device (SQUID) magnetometer. Heat capacity was measured from 1.9 to 40 K on a PPMS by mounting pressed pellets of the samples on a sapphire platform with Apiezon N grease. The electronic structure was calculated using WIEN2k.[16] The structures used to perform the calculations were based on full structural optimization results from VASP [17], starting with the experimentally determined structures. (Detailed calculations using LMTO methods for $Hf_5Sb_{2.5}M_{0.5}$ for M= 3$d$ transition metals have previously been described.[9])

The refined crystal structure for $Hf_5Sb_{2.5}Ru_{0.5}$ is shown in Figure 1(Upper). This compound, and the others studied here, adopts the tetragonal $W_5Si_3$-type structure (space group $I4/mcm$, Pearson Symbol $tP32$). Hf atoms are located at the $4b$ and $16k$ sites corresponding to the W sites in $W_5Si_3$ and Sb atoms fully occupy the $8h$ site. We have tested for the possibility of Sb/Hf mixed occupancy in the refinements and did not observe it to be significant. The $4a$ sites are occupied by a nearly 1:1 mixture of Sb and Ru. The figure shows that the structure consists of $Hf_8$ square antiprisms sharing their square faces to create columns along $c$, with linear chains of randomly mixed Ru/Sb atoms, in a nearly 1:1 ratio, imbedded inside them. The columns of square antiprisms are surrounded by Sb-Sb zigzag chains. The variability of Ru content allowed in the compound involves changes in the ratio of Ru to Sb in the linear chains. Detailed crystallographic data is presented in the supplementary information. Inspection of the data in Tables S2 and S3 indicates that the equivalent isotropic displacement parameter of the $8h$ (Sb3) sites is lowest among the four positions in the asymmetric unit; our refinements show that this is not due to the presence of Hf/Sb mixing, and we note that relatively smaller displacement parameters are consistently observed at the $8h$ sites for $Hf_5Sb_{3-x}M_x$ ($M$ = V, Mn, Fe and Ni).[9]

The temperature ($T$) dependent electrical resistivity for $Hf_5Sb_{2.5}Ru_{0.5}$ between 1.9 and 300 K is shown in Figure 2($a$). The resistivity undergoes a sudden drop to zero at 3.2 K, characteristic of superconductivity. In correspondence with $\rho(T)$, the magnetic susceptibility ($\chi_{mol}(T)$)–starts to decrease at 3.2 K and shows large negative values, characteristic of a fully superconducting sample. The zero resistivity and the large diamagnetic susceptibility indicate that $Hf_5Sb_{2.54(4)}Ru_{0.46}$ becomes a bulk superconductor at 3.2 K. Critically, only the Ru doped compound shows the presence of superconductivity; $Hf_5Sb_{2.5}M_{0.5}$ (M=Mo, Rh, Pd, Re, Ir and Pt) show no superconductivity. For example, $Hf_5Sb_{2.5}Mo_{0.5}$ and $Hf_5Sb_{2.5}Rh_{0.5}$ show only weak core-diamagnetism-dominated magnetic susceptibilities (Figure 2($a$), insert). To prove that the observed superconductivity is intrinsic to the $Hf_5Sb_{2.5}Ru_{0.5}$ compound, the superconducting transition was characterized further, through specific heat measurements. Specific heat measurements are a reliable indication of the presence of bulk superconductivity when combined with resistivity and susceptibility measurements due to the change in bulk thermodynamic properties at the superconducting transition. The specific heat for $Hf_5Sb_{2.5}Ru_{0.5}$ in the temperature range of 1.9 K to 40 K is presented in Figure 2($b$). The main panel shows the temperature dependence of the zero-field and field-cooled specific heat $C_p/T$. The good quality of the sample and the bulk nature of the superconductivity are strongly supported by the presence of a large anomaly in the specific heat at $T_c$= 3.2 ~ 3.3 K, in excellent agreement with the $T_c$ determined by ρ(T) and $χ_{mol}$(T). The electronic contribution to the specific heat, γ, measured in a field of 5 T to suppress the superconductivity (lower inset to Figure 2(*b*)), is 21.92 mJ/mol-K$^2$. (The data fitted using the formula $C_p = γT + βT^3$, in which γ and β are the electronic and lattice contributions to the specific heat, respectively.) The value of the specific heat jump at $T_c$, determined by the equal area method (upper inset Figure 2(*b*)), is consistent with that expected from a weak-coupling BCS superconductor; $ΔC_{el}/γT_c$ per mole $Hf_5Sb_{2.5}Ru_{0.5}$ in the pure sample = 1.46. This ratio is within error of the BCS superconductivity weak coupling value of 1.43, and is in the range observed for many superconductors.[18] As an added check, we tested pure HfRu (The impurity that is present when the Ru content of a sample exceeds the solubility limit of the $Hf_5Sb_{3-x}Ru_x$ phase.) down to 1.78 K and found that it is not superconducting; that compound therefore could not give rise to the observed specific heat feature. Thus the observed superconductivity originates from $Hf_5Sb_{2.5}Ru_{0.5}$. The figure (lower inset) also shows that the Ru-Sb compound has a significantly larger electronic contribution to the specific heat (γ) than the two comparison materials. The data indicates γ values of 21.9, 13.9 and 15.9 mJ/mol-K$^2$ for the Ru-Sb, Mo-Sb and Rh-Sb compounds respectively. This parameter is a reflection of the density of electronic states at the Fermi energy and its renormalization due to electron-phonon coupling[19,20], and in the current case the large value for the Ru-Sb couple is an indication of the propensity for superconductivity.

To gain further insight into the uniqueness of $Hf_5Sb_{2.5}Ru_{0.5}$, we investigated the electronic density of states (DOS) and band structures of the hypothetical compounds "$Hf_{10}Sb_5M$" (i.e. $Hf_5Sb_{2.5}M_{0.5}$; $M$= Fe, Ru, Mo and Rh) in space group *I*422, which allows the *M*-Sb chains to be modelled as *M* mixed with Sb in an alternating 1:1 ratio on what is the 4*a* site in the $W_5Si_3$-type structure. This provides insight into the significance of the transition metal states near the Fermi energy ($E_F$) and the differences between 3*d* and 4*d* *M* systems. We first focus on the comparison between $Hf_{10}Sb_5Fe$ (3*d*) and $Hf_{10}Sb_5Ru$ (4*d*). The calculated DOS curves and band structures for these compounds are illustrated in Figure 3(*a*), which emphasizes contributions in the range from -1eV to +1eV of $E_F$. The significant DOS at $E_F$ is consistent with the metallic properties of $Hf_{10}Sb_5Fe$ and $Hf_{10}Sb_5Ru$, and analysis of the orbitals contributing to the bands at $E_F$ shows the dominance of the transition metal *d* states in this energy regime. Further, a saddle point is found near the N point in the Brillouin zone. The existence of such saddle points near $E_F$, called "van Hove singularities", leads to the presence of peaks in the electronic density of states and is considered to be important in yielding superconductivity in a variety of superconductors, including oxo-cuprates, dichalcogenides and even niobium.[21] We hypothesize that this peak in the DOS remains present even for a disordered or short-range ordered Ru-Sb chain structure. To test this hypothesis, we also calculated the electronic density of states for a chain arrangement of the type -Sb-Sb-Ru-Ru-, which may be present in small proportion in the material. The resulting DOS is shown in Figure S2. The peak in the DOS remains present. Thus these initial electronic structure calculations confirm the appropriateness of this structure type for the test of our hypothesis – they indicate both the significant influence of the transition metal (*M*) electronic states at $E_F$, even though the *M* elements are a minor elemental constituent of the phase, and the presence of a peak in the electronic density of states, suggesting that superconductivity may be found in these compounds. The calculations suggest that the 3*d* and the 4*d* transition metal variants possess almost the same electronic picture.

Taking the argument one step further, we compare in Figure 3(*b*) the calculated band structures of $Hf_{10}Sb_5M$ (*M* = Mo, Ru and Rh) in this structure type to obtain information about why 4*d* transition metals other than Ru did not induce superconductivity. The band structure calculations show that $E_F$ for the Ru-based compound locates close to the saddle point at N. Significantly, the saddle point at N is sensitive to the *M* element present – it is split into two bands for both the Mo and Rh cases, that is to say, a 'fingerprint' characteristic for superconductivity is not present as robustly in the Mo and Rh cases as it is for the Ru case. Our motivation for making the $Hf_{4.5}Y_{0.5}Sb_{2.5}Rh_{0.5}$ (i.e. $Hf_9YSb_5Rh$) compound is based on this observation. Removing one electron from $Hf_{10}Sb_5Rh$ puts $E_F$ at the right position in the band structure to yield a compound that is structurally and electronically equivalent to $Hf_{10}Sb_5Ru$. We find, however, that $Hf_9YSb_5Rh$ is not superconducting.

In conclusion, we have shown that the $Hf_5Sb_{3-x}M_x$ family of compounds allows us to examine the hypothesis that chemical identity is important for superconductivity in intermetallic phases. We therefore propose that like Cu-O and Fe-As in other types of compounds, Ru and Sb represent a critical element pair for superconductivity in intermetallic compounds. This conclusion supplements the long standing belief, based on a "rigid band" picture for intermetallics, that crystal structure and electron count (which, in a physics-based picture determine the Fermi surface) are the primary crystal-chemical requirements for the superconducting state in intermetallics, by adding another component for consideration. The work described here shows in general that when searching for new superconductors, even when given favorable electron counts and crystal structures, different but seemingly equivalent elemental constituents should be tested, simply because not all atoms are the same, even in superconductors.

**Acknowledgement**

The chemical synthesis, single crystal diffraction and electronic structure calculations were supported by the Department of Energy, grant DE-FG02-98ER45706. The measurement and interpretation of the physical properties were supported by the Gordon and Betty Moore Foundation's EPiQS Initiative through Grant GBMF4412. M.B.N. gratefully acknowledges the Danish Ministry of Higher Education and Science for a travel scholarship for its support of his visit to Princeton.

**Supporting Information**

The single crystal data, atomic coordinates and anisotropic thermal parameters of $Hf_5Sb_{2.54(4)}Ru_{0.46}$, powder XRD diffraction patterns of $Hf_5Sb_{3-x}M_x$ ($M$ = Mo, Ru and Rh); Density of States of $Hf_{10}Sb_5Ru$ with -Ru-Sb- linear chain and -Ru-Sb-Sb-Ru-Ru-Sb- linear chain, CIF and check-cif files of single crystal data of $Hf_5Sb_{2.54(4)}Ru_{0.46}$. This information is available free of charge via the Internet at http://pubs.acs.org.

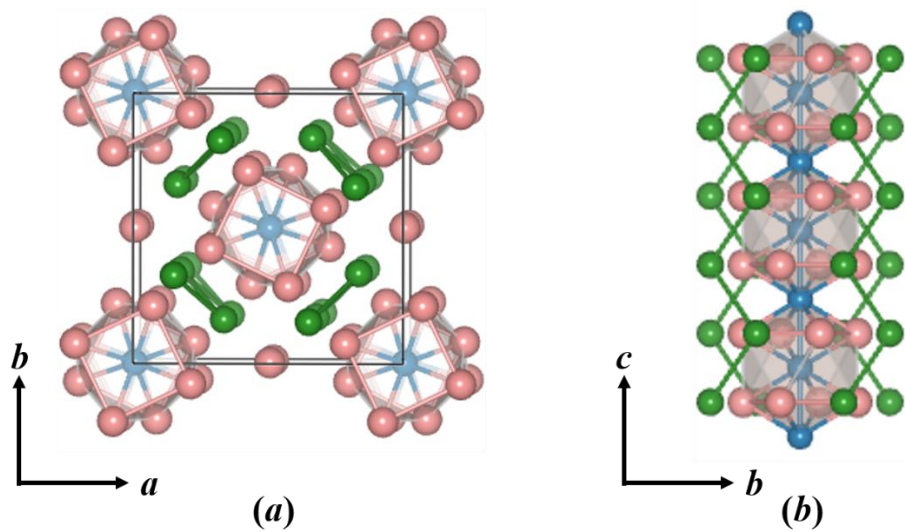

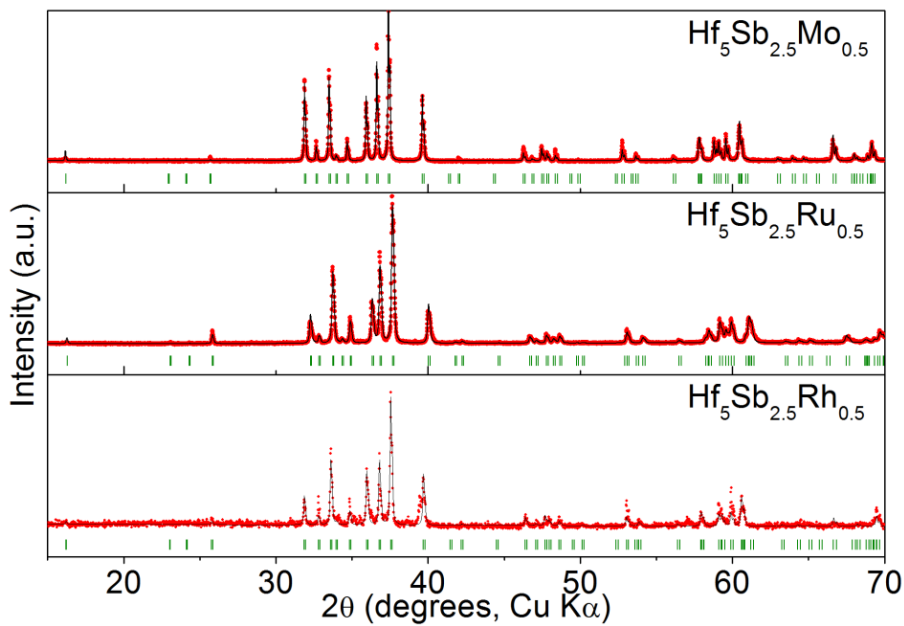

**Figure 1. (Upper) The crystal structure of $Hf_5Sb_{2.54(4)}Ru_{0.46}$ in the $W_5Si_3$-type structure.** (*a*) a (001) view emphasizing the Hf square antiprisms around the 1:1 Ru/Sb chains. (Green: Sb; blue: Ru/Sb mixed chains; purple: Hf) (*b*) a (100) "side view" of a column of Hf with the imbedded Ru-Sb chain. The Sb atoms surrounding the Hf column are also shown. **(Bottom) Powder x-ray diffraction data showing the pure phases of $W_5Si_3$-type $Hf_5Sb_{3-x}M_x$ ($M$ = Mo, Ru and Rh).** Red solid line shows the corresponding Rietveld fitting.

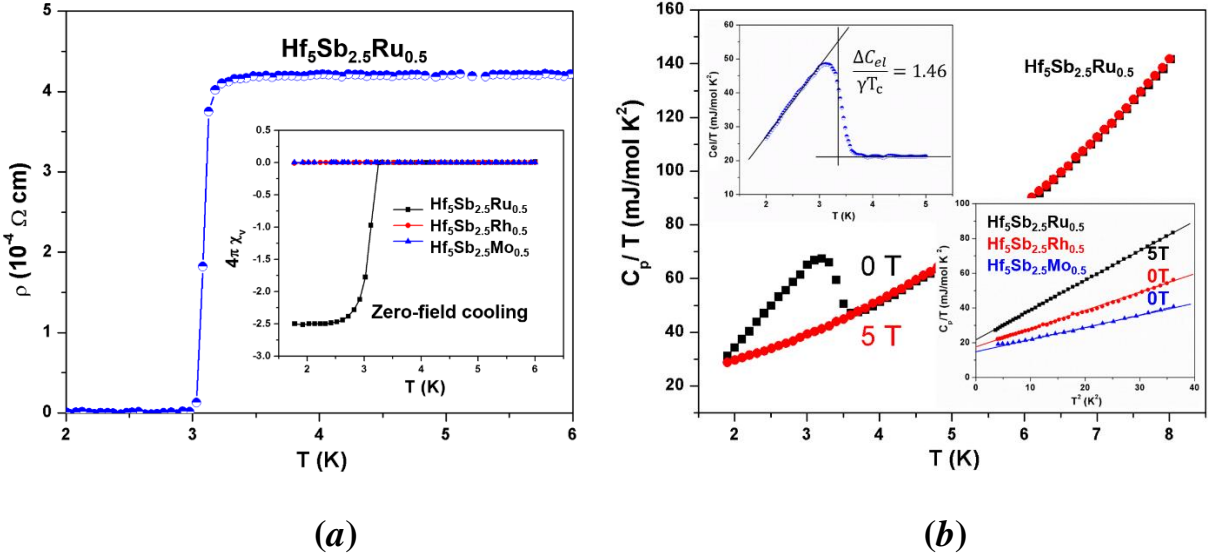

**Figure 2. Physical properties measurements.** (*a*) **The electrical and magnetic properties of the Hf$_5$Sb$_{2.5}$Ru$_{0.5}$ superconductor.** (*Main panel*) The temperature dependence of the electrical resistivity of Hf$_5$Sb$_{2.5}$Ru$_{0.5}$ in 0 applied magnetic field showing a close-up of the superconducting transition. (*Insert*) The temperature dependence of the magnetic susceptibility for Hf$_5$Sb$_{2.5}$M$_{0.5}$ (M=Mo, Ru and Rh) between 1.8K and 6K in an applied field of 10 Oe after zero-field cooling. (*b*) **Specific heat characterization of the superconducting transition of Hf$_5$Sb$_{2.5}$Ru$_{0.5}$.** (*Main panel*) Temperature dependence of the specific heat C$_p$ of a Hf$_5$Sb$_{2.5}$Ru$_{0.5}$ sample measured with ($\mu_0$H = 5T) and without a magnetic field, presented in the form of C$_p$/T vs. T. (*Insert, Lower*) C$_p$/T vs. T of Hf$_5$Sb$_{2.5}$M$_{0.5}$ (M= Mo, Ru and Rh) at applied fields of 0 (M=Mo and Rh) and 5T (to suppress the superconductivity) fit to the form C$_p$/T = $\gamma$ + $\beta$T$^2$; $\gamma$ is the electronic contribution to the specific heat and $\beta$T$^2$ is the contribution of lattice vibrations. (*Insert, Upper*) The low temperature electronic heat capacity, as C$_{el}$/T vs. T, in the temperature range 2.0-5.0 K; this is the "equal area construction" employed to determine the ratio of the change in entropy at the superconducting transition to the electronic specific heat ($\gamma$). C$_{el}$ is determined by Cp ($\mu_0$H =0) – $\beta$T$^2$, where the latter part is the phonon part of the specific heat (see lower insert). The data show that Hf$_5$Sb$_{2.5}$Ru$_{0.5}$ is a high quality weak coupling BCS superconductor.

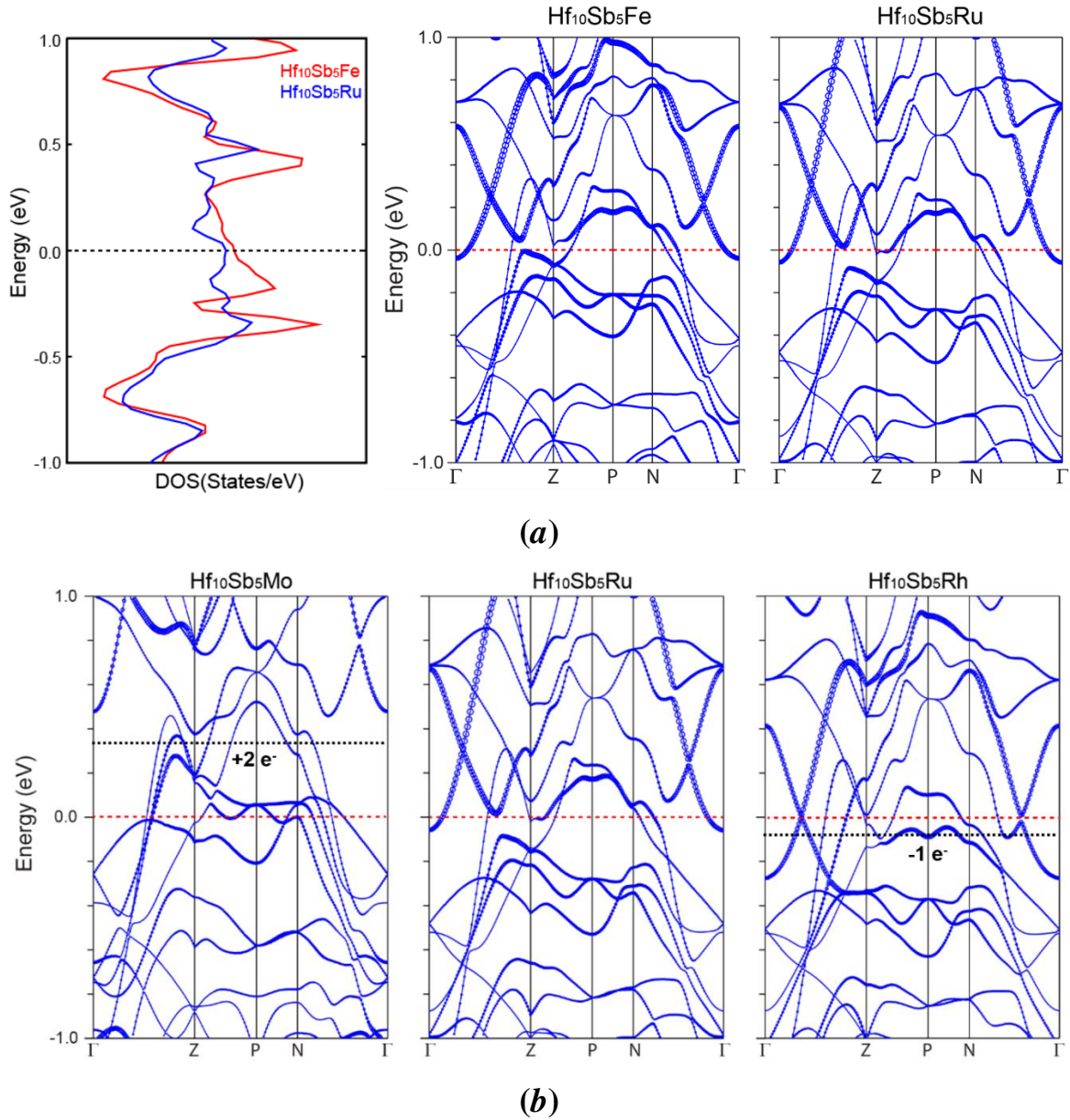

**Figure 3.** (*a*) **Results of the Electronic structure calculations for the hypothetical model compounds "Hf$_{10}$Sb$_5$Ru" and "Hf$_{10}$Sb$_5$Fe".** Total DOS curves and band structure curves are obtained from non-spin-polarized LDA calculations. (*b*) **Results of the band structure calculations for hypothetical model compounds "Hf$_{10}$Sb$_5$M" (M=Mo, Ru and Rh) obtained from non-spin-polarized LDA calculations.** The Ru calculation is a good representation of the superconducting phase. The red dashed lines show the Fermi levels for each compound, and the black dashed lines show where the Fermi energy would fall when the appropriate number of electrons are added (Mo) or subtracted (Rh) to make these compounds have the same electron count as the Ru variant.

**Table of Content**

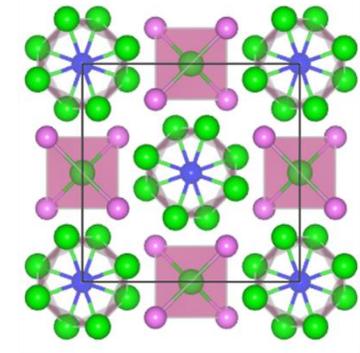

**New Superconductor**
$Hf_5Sb_{3-x}Ru_x$

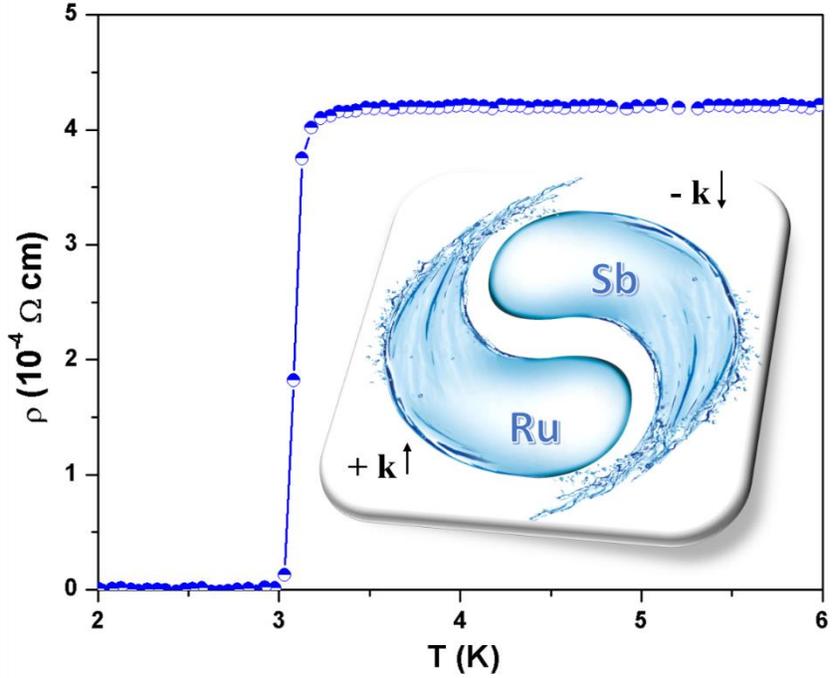

We present the way to searching for a new superconductor in $W_5Si_3$-type $Hf_5Sb_{3-x}Ru_x$ solid solution by coupling theoretical efforts and experimental approach. Our study leads us to propose that, like Cu-O and Fe-As in high temperature superconductors, Ru and Sb may be a critical element pair for superconductivity in intermetallic phases.